\pdfoutput=1
\documentclass[aip,reprint,apl]{revtex4-1}
\usepackage{graphicx}
\usepackage{amsmath}
\usepackage{amssymb}
\usepackage{booktabs}
\usepackage{array}
\usepackage{xcolor}
\usepackage{acronym}
\usepackage{pifont}
\usepackage[normalem]{ulem}

\begin{document}
\title{Terahertz electron paramagnetic resonance generalized spectroscopic ellipsometry: The magnetic response of the nitrogen defect in 4H-SiC}

\author{Mathias Schubert}
\affiliation{Department of Electrical and Computer Engineering, University of Nebraska-Lincoln, Lincoln, NE 68588, USA}
\affiliation{Terahertz Materials Analysis Center and Center for III-N Technology, C3NiT -- Janz\`{e}n, Department of Physics, Chemistry and Biology (IFM), Link\"{o}ping University, 58183 Link\"{o}ping, Sweden}
\email{schubert@engr.unl.edu}
\homepage{http://ellipsometry.unl.edu}
\author{Sean Knight}
\affiliation{Terahertz Materials Analysis Center and Center for III-N Technology, C3NiT -- Janz\`{e}n, Department of Physics, Chemistry and Biology (IFM), Link\"{o}ping University, 58183 Link\"{o}ping, Sweden}
\author{Steffen Richter}
\affiliation{Terahertz Materials Analysis Center and Center for III-N Technology, C3NiT -- Janz\`{e}n, Department of Physics, Chemistry and Biology (IFM), Link\"{o}ping University, 58183 Link\"{o}ping, Sweden}
\author{Philipp K\"{u}hne}
\affiliation{Terahertz Materials Analysis Center and Center for III-N Technology, C3NiT -- Janz\`{e}n, Department of Physics, Chemistry and Biology (IFM), Link\"{o}ping University, 58183 Link\"{o}ping, Sweden}
\author{Vallery Stanishev}
\affiliation{Terahertz Materials Analysis Center and Center for III-N Technology, C3NiT -- Janz\`{e}n, Department of Physics, Chemistry and Biology (IFM), Link\"{o}ping University, 58183 Link\"{o}ping, Sweden}
\author{Alexander Ruder}
\affiliation{Department of Electrical and Computer Engineering, University of Nebraska-Lincoln, Lincoln, NE 68588, USA}
\author{Megan Stokey}
\affiliation{Department of Electrical and Computer Engineering, University of Nebraska-Lincoln, Lincoln, NE 68588, USA}
\author{Rafa\l{} Korlacki}
\affiliation{Department of Electrical and Computer Engineering, University of Nebraska-Lincoln, Lincoln, NE 68588, USA}
\author{Klaus Irmscher}
\affiliation{Leibniz-Institut f\"{u}r Kristallz\"{u}chtung, 12489 Berlin, Germany}
\author{Petr Neugebauer}
\affiliation{CEITEC – Central European Institute of Technology, Brno University of Technology, Brno, Czech Republic}
\author{Vanya Darakchieva}
\affiliation{Terahertz Materials Analysis Center and Center for III-N Technology, C3NiT -- Janz\`{e}n, Department of Physics, Chemistry and Biology (IFM), Link\"{o}ping University, 58183 Link\"{o}ping, Sweden}
\affiliation{Solid State Physics and NanoLund, Lund University, P. O. Box 118, S-221 00 Lund, Sweden}
\email{vanya.darakchieva@liu.se}
\homepage{https://c3nit.se/; https://liu.se/en/research/themac}

\date{\today}

\begin{abstract}
We report on terahertz (THz) electron paramagnetic resonance generalized spectroscopic ellipsometry (THz-EPR-GSE). Measurements of the field and frequency dependencies of the magnetic response due to the spin transitions associated with the nitrogen defect in 4H-SiC are shown as an example. THz-EPR-GSE dispenses with the need of a cavity, permits independently scanning field and frequency parameters, and does not require field or frequency modulation. We investigate spin transitions of hexagonal ($h$) and cubic ($k$) coordinated nitrogen including coupling with its nuclear spin (I=1), and we propose a model approach for the magnetic susceptibility to account for the spin transitions. From the THz-EPR-GSE measurements we can fully determine the polarization properties of the spin transitions and we obtain $g$ and hyperfine splitting parameters using magnetic field and frequency dependent Lorentzian oscillator lineshape functions.  We propose frequency-scanning THz-EPR-GSE as a new and versatile method to study properties of spins in solid state materials.
\end{abstract}

\maketitle


Electron paramagnetic resonance (EPR) is ubiquitous in science.\cite{Poole1983} Traditional EPR instruments operate in the lower Gigahertz (GHz) range limited to one or a few frequencies only.\cite{doi:10.1021/ja003707u} The possibility to access electron spin dynamics at much higher, i.e., Terahertz (THz) frequencies is attractive. Large energies permit better understanding of spin dynamics in single-molecule magnets, for example,\cite{C7CP07443C,LAGUTA2018138} and allow investigation of systems with large zero-field splitting such as in transition metal complexes,\cite{doi:10.1021/ja00139a021,HASSAN2000300} or in ultrawide-bandgap semiconductors, e.g., SiC\cite{ PhysRevB.103.245203, doi:10.1063/1.4866331, SonAMR2010} and group-III nitrides\cite{ ZvanutJEM2019, Sunay_2019} for quantum technologies, or the recently emerging monoclinic gallium oxide for high voltage electronic applications.\cite{doi:10.1063/1.4990454,doi:10.1063/1.5053158,doi:10.1063/1.4972040,doi:10.1063/5.0002763, doi:10.1063/1.5133051,doi:10.1063/1.5127651,doi:10.1063/1.5081825,doi:10.1063/5.0012579} To bring spin resonances into the THz range, superconducting magnets reaching large magnetic fields are necessary. The emergence of closed-cycle dry magnet systems and novel superconducting materials have made 20~T field-flattened single-solenoid EPR magnets possible.\cite{WOS:000535356200018} In the THz spectral range one can employ optical methods such as reflectance and transmittance measurements using free space plane wave propagation. Advantages of large-field frequency-scanning far-field THz EPR are manifold. The need for a fixed cavity and fixed frequency tied to the cavity is dispensed with. Furthermore, frequency scanning of spin systems permits to easily verify, for example, whether a given signature is caused by hyper fine structure splitting ($hfs$) or due to Zeeman splitting of multiple different species since the $hfs$ splitting is independent of static magnetic field strength. Furthermore, because the spin susceptibility is proportional to the population difference between the Zeeman split spin levels, the susceptibility decrease with increasing temperature from its maximum at T~=~0~K, mitigated due to the much higher Zeeman splitting at the more than ten fold higher static magnetic fields applied.\cite{Poole1983,WeilBolton2007} The energy resolution of the EPR signatures improves with higher frequencies at constant bandwidth, and the sensitivity to spin densities increases with frequency $\omega$ by $(\omega/\omega_0)^{3}$, for example between 10~GHz and 0.225~THz by approximately four orders of magnitude (Ref.~\onlinecite{WeilBolton2007}, Appendix~F.3.3.). Furthermore, samples with large surface area can be investigated.

Conventional EPR measures the loss of an electromagnetic wave upon the spin transition absorption process within a specimen that is placed within the maximum of a near-field pattern of a known resonant microwave cavity. Background and signal instabilities are typically overcome by detecting  absorbance difference signals augmenting small magnetic field oscillations. In recent years, V-band (72 GHz), W-band (95 GHz), and higher band systems have become available.\cite{SavitskyPhotosynthRes2009} These, however, are often based on absorbance measurements and use resonance cavities with reduced dimensions requiring also smaller sample size.\cite{Poole1983,WeilBolton2007} Millimeter waveguides suffer from high insertion losses at off resonance frequencies and hamper frequency-scanning EPR. A waveguide-based normal-incidence THz high field (15~T) frequency domain (85--720~GHz) EPR spectrometer with co- and cross-polar detection of differential absorbance was presented by Neugebauer~\textit{et al.}\cite{C7CP07443C,Bloos_2019}

The magnetic polarization causes dispersion in the real and imaginary parts of the magnetic permeability, $\mu$, which modify the complex index of refraction, $n=\sqrt{\mu\varepsilon}$, where $\varepsilon$ is the dielectric permittivity. It is well known from thin film optics that small modifications in both real and imaginary parts of the complex index of refraction can precisely be monitored by analysis of the state of polarization of reflected or transmitted light. The preeminent method to investigate the optical properties of solid state materials and thin films is spectroscopic ellipsometry (SE).\cite{Fujiwara_2007} Measurement of properties from anisotropic materials requires generalized SE (GSE).\cite{SchubertJOSAA13_1996,SchubertADP15_2006} GSE measures the so called Mueller matrix elements upon analysis of changes in polarization of a polarized plane wave reflected or transmitted from a sample. The Mueller matrix contains the most complete polarization information of any given sample.\cite{MuellerMIT1943} Measurements of magneto-optical (MO) properties of samples using GSE (MO-GSE) are widely used to study, e.g., free charge carrier properties and magnetic domain dynamics.\cite{WOS:000170203300002,WOS:000175575100159,WOS:000180643200016} In the THz spectral range, MO-GSE measurements are very sensitive to free charge carrier properties, i.e., the Optical Hall effect (OHE).\cite{Schubert:16} No EPR Mueller matrix measurements exist.

In this letter, we report on THz-EPR generalized spectroscopic ellipsometry (THz-EPR-GSE) measurements of the field and frequency dependencies of the magnetic response of the spin transitions associated with the nitrogen defect in 4H-SiC, as example. We demonstrate frequency-scanning THz-EPR-GSE as a new and versatile method to study properties of spins in solid state materials. We investigate spins in hexagonal ($h$) and cubic ($k$) coordinated nitrogen and coupling with the nuclear nitrogen spin (I=1), and propose a model approach for the spin transitions observed here in the magnetic susceptibility. We directly determine real and imaginary parts of the magnetic susceptibility and thereby the polarization properties of the spin transitions. We emphasize that our technique permits independently scanning frequency and field parameters, dispenses with the need for cavities, and does not require field or frequency modulation. We propose THz-EPR-GSE as a new approach to determine the complex magnetic response functions of EPR active materials. 

Silicon carbide (SiC) is a well investigated wideband gap semiconductor material. EPR signatures of defects in SiC are also well studied, and SiC can serve as excellent standard for the THz-EPR-GSE method presented here.\cite{GreulichWeber1997,SonAMR2010} In 4H-SiC N can be incorporated into hexagonal ($h$) and quasi-cubic ($k$) sites. Both have slightly anisotropic $g$-factors for magnetic field direction parallel ($||$) and perpendicular ($\perp$) to the lattice $c$ axis ($h$: $g_{\mathrm{||}}$~=~2.0055, $g_{\perp}$~=~2.0043; $k$: $g_{\mathrm{||}}$~=~2.0043, $g_{\perp}$~=~2.0013). The spins ($m_s=\pm\frac{1}{2}$) couple with the nuclear spin of $^{14}$N ($I$ = 1) and thus form triplets through $hfs$ interaction. The $h$ site reveals a small $hfs$ coupling constant ($\alpha_{h,hfs}$~= 2.9~MHz or 0.10~mT) while the $k$ site shows a stronger $hf$ splitting ($\alpha_{k,hfs}$~= 50.97~MHz or 1.8~mT).\cite{SonAMR2010, doi:10.1063/1.4866331, PhysRevB.103.245203, PhysRevB.70.193207} 


We model the magneto-optic anisotropy due to EPR transitions assuming that level transitions with $\Delta m_s = \pm 1$ correspond to absorption by right (-, RCP) or left (+, LCP) handed circularly polarized electromagnetic plane waves for propagation parallel to the static magnetic field ($\mathbf{B}=B\mathbf{\hat{e}}$) orientation ($\mathbf{\hat{e}}$). To begin with, $\mathbf{B}$ is parallel $z$. We seek contributions to the magnetic polarization phasor, $\mathbf{M}$, and a pair of magnetic response functions, $\sigma_{\mathrm{\pm}}$, may depend on photon energy, $\hbar\omega$, and magnetic field, $\mathbf{B}$.\footnote{This derivation is analogous to the consideration for magneto-optic dielectric anisotropy due to, for example, Landau level transitions in two-dimensional charge carrier systems discussed in detail in Ref.~\onlinecite{Schubert:16}.} $\mathbf{M}_C = diag(\sigma_{\mathrm{+}},\sigma_{\mathrm{-}},0)\mathbf{H}_C$, where $diag$ indicates the diagonal matrix, $\mathbf{M}_C =(\sigma_{\mathrm{+}}H_+,\sigma_{\mathrm{-}}H_-,0)$, and $\mathbf{H_{C}} =(H_+,H_-,H_z)= (\frac{1}{\sqrt{2}}[H_x+iH_y],\frac{1}{\sqrt{2}}[H_x-iH_y],H_z)$. After transformation into the $(x,y,z)$ laboratory system, $\mathbf{M} = \{M_x,M_y,M_z\}=\chi_{m}\{H_x,H_y,H_z\}$, with the magnetic susceptibility tensor, $\chi_{m}$ 

\begin{equation}
\chi_m\left(\pm B\right)=\frac{1}{2}
\begin{pmatrix}
\left[\sigma_{+}+\sigma_{-}\right] & \mp i\left[\sigma_{+}-\sigma_{-}\right] & 0\\
\pm i \left[\sigma_{+}-\sigma_{-}\right] & \left[\sigma_{+}+\sigma_{-}\right] & 0\\
0&0&0
   \end{pmatrix},\label{eq:EPRtensor}
\end{equation}

\noindent where $\sigma_{+(-)}$ is the magnetic response function for LCP (RCP), and $\{H_x,H_y,H_z\}$ is the magnetic field phasor of the electromagnetic wave. The magnetic permeability is $\mu=\mu_0(\mathbf{I}+\chi_m)$, where $\mathbf{I}$ is the unit matrix and $\mu_0=4\pi\times 10^{-7}$H/m is the vacuum permeability. The meaning of RCP/LCP reverses upon reversal of the magnetic field, hence, the off diagonal elements switch sign. The absorption of electromagnetic waves in a spin transition requires transfer of angular momentum, hence, only one type of circular polarization is absorbed ($\sigma)$, while the magnetic function for the other polarization is zero

\begin{equation}
\chi_m\left(\pm B\right)=\frac{1}{2}
\begin{pmatrix}
\sigma & \mp i\sigma & 0\\
\pm i \sigma & \sigma& 0\\
0&0&0
   \end{pmatrix}.\label{eq:EPRtensorSiC}
\end{equation}

\noindent We render $\sigma$ with sums of Lorentz functions, $\sigma=\sum \sigma_{j}$, which represent the spin polarizability due to transition between $\mathbf{B}$ field dependent levels with energy $E_1$ and $E_2$, with photon energy, $\hbar\omega =\Delta E(\mathbf{B}) = E_2-E_1$

\begin{equation}
    \sigma_{j}=\frac{A^2}{\left(\Delta E_{j}(\mathbf{B})\right)^2-(\hbar\omega)^2-i\hbar\omega\hbar\gamma_{j}},\label{eq:lorentz}
\end{equation}

\noindent and where we have included a line broadening parameter, $\gamma$, and whose role is to render effects of life time and transition energy broadening, e.g., by magnetic field inhomogeneity. For coupling with nuclear spin $I=1$, the electron spin transitions split into a triplet ($j=1,\dots,3$). The transition energy is dependent on magnitude and direction of the magnetic field

\begin{equation}
    \Delta E(\mathbf{B})=\mu_B(\mathbf{\hat{e}}\mathbf{g}\mathbf{\hat{e}}^{T}B+\Delta m_I\alpha_{hfs}),\mbox{}\Delta m_I=-1,0,1,
\end{equation}

\noindent where $\mu_B$ is the Bohr magneton, $\alpha_{hfs}$ is the $hfs$ splitting constant, $T$ denotes the transpose of a vector, and $\mathbf{g}$ is the fully symmetric anisotropic $g$-tensor, $\mathbf{g}=diag\{g_{xx},g_{yy},g_{zz}\}$.We obtain the value of $\Delta E(\mathbf{B})$ by rotating the magnetic field to its orientation within a given material using Euler rotations $\mathbf{A}(\theta,\varphi)$. Euler angles $\theta$ and $\varphi$ describe inclination from the $z$ axis and in-plane azimuth of the projection of the magnetic field onto the $x-y$ plane. In our ellipsometer system, the sample surface is the $x-y$ plane, the plane of incidence is the $x-z$ plane. For (0001) 4H-SiC the surface is oriented with the $c$ axis parallel to the sample normal, where $g_{zz} = g_{||}$, and  $g_{xx,yy} = g_{\perp}$. 

Calculated Mueller matrix data are obtained using a two-layer model approach, where an anisotropic layer stack is placed atop a mirror (the metal substrate holder) followed by an ambient gap (layer 1; index $n$=1, thickness 18.4692~$\mu$m), the SiC substrate (layer 2; thickness 505.414~$\mu$m) and normal ambient (index $n$=1). The layer rendering the THz optical properties of the SiC substrate includes the uniaxial optical character by considering ordinary ($n_{\perp}=3.197$) and extraordinary ($n_{||}=3.314$) refractive indices obtained from zero field measurements here, close to results obtained by Naftalny~\textit{et. al.}\cite{Naftaly:16} Our previously described matrix formalism is used for calculations, which also permits for anisotropy in the magnetic permeability.\cite{PhysRevB.53.4265} The optical properties of the SiC sample are now rendered by two tensors, $\varepsilon$ and $\mu$. Thickness data are determined from model analysis of zero-field data as described previously.\cite{KnightRSI2020}

\begin{figure}[ht]
\centering
\includegraphics[width=1.0\linewidth]{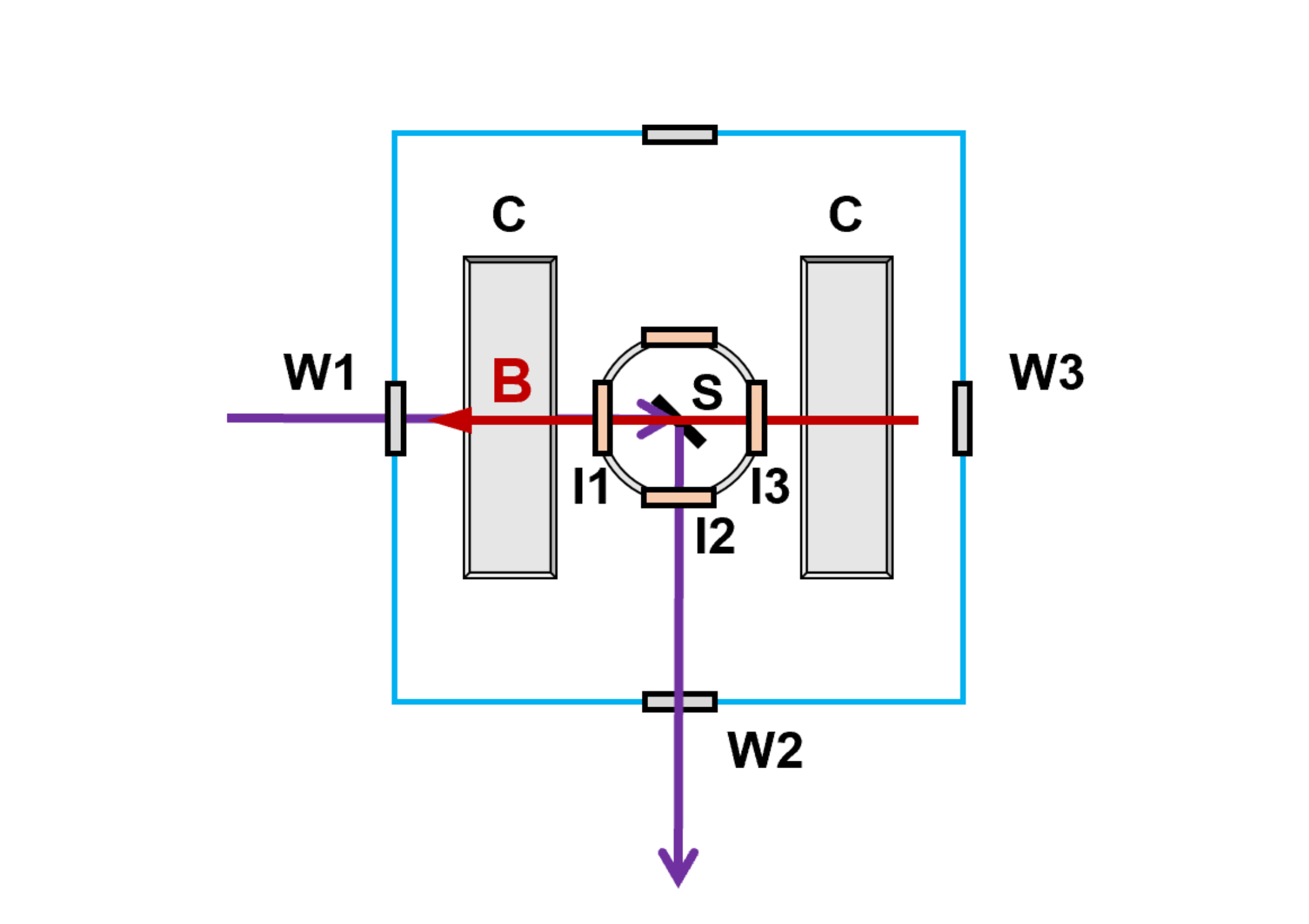}
\caption{Principle drawing of the THz-EPR-E setup. A split-coil (C,C) superconducting magnet produces field B with direction along the incident beam. External (W1, W2, W3; polymer) and internal (I1, I2, I3; diamond) windows separate coils and sample (S) inside the cryostat. A polarization state generator and polarization state detector (not shown) create and analyze state of polarization before and after the sample, respectively.}
\label{fig:THzEPRE}
\end{figure}


We investigate a nitrogen doped ($N_{\mathrm{C}}=1\times$10$^{18}$~cm$^{-3}$) (0001)-oriented 4H-SiC substrate. The sample is placed within a split-coil superconducting magnet at 10~K.\cite{WOS:000431446900001} The magnetic field is directed at 45$^{\circ}$ towards the $c$ axis, and parallel to the incident THz beam (Fig.~\ref{fig:THzEPRE}). Hence, $\theta=45^{\circ}$ and $\varphi=-90^{\circ}$. Prior to measurements, we calibrated the magneto-optical cryostat magnetic field sensor using a thick film of 1,1-diphenyl-2-picryl-hydrazyl (DPPH) bound by cyanoacrylate (super glue) onto a 400~$\mu$m thick polypropylene substrate, and we monitored the spin transition ($g = 2.0036$) as a function of the current control settings at 130~K temperature.\cite{KRZYSTEK1997207} For our magnet, we found that for the true magnetic field amplitude ($B$) a small linear correction must be applied to the set value ($B_{\mathrm{set}}$) of the magnetic field ($B=1.0046B_{\mathrm{set}}$). The $g$-factors for our setup's field orientation can be calculated according to ($j$="$h$,$k$")\cite{GreulichWeber1997}

\begin{equation}
    g_j = g_{||,j}\sin^2\theta + g_{\perp,j}\cos^2\theta.
\end{equation}

\noindent At $\theta$~=~45$^{\circ}$, the $g$ factors are 2.0028 ($k$) and 2.00325 ($h$), with a splitting of 0.00045.

We have measured the Mueller matrix elements in our THz-OHE instrument as described previously.\cite{WOS:000431446900001} Data were obtained at multiple fixed frequencies as a function of the magnetic field (negative field scans: at 120, 130, 137.7, 138, 150, and 160~GHz; positive field scans: at 120, 130, 134, 136.6, 137.7, 138, 139.5, 150, 160, and 164~GHz), where the magnetic field was progressed in steps of $0.2\dots0.5$~mT for both positive and negative field direction scans, and at fixed positive and negative magnetic fields ($\pm$~4.8628~T, $\pm$~4.9338~T, $\pm$~4.8984~T), where the frequency was increased in steps of 10~MHz. Scans were performed in frequency and field regions where the spin resonance was anticipated. In our current setup, only the upper 3$\times$3 block of the Mueller matrix is measured, and all elements are normalized by $M_{11}$. For all data, we subtract the zero field value, and show and discuss below magnetic field induced differences in the Mueller matrix elements.

\begin{figure}[ht]
\centering
\includegraphics[width=1.00\linewidth]{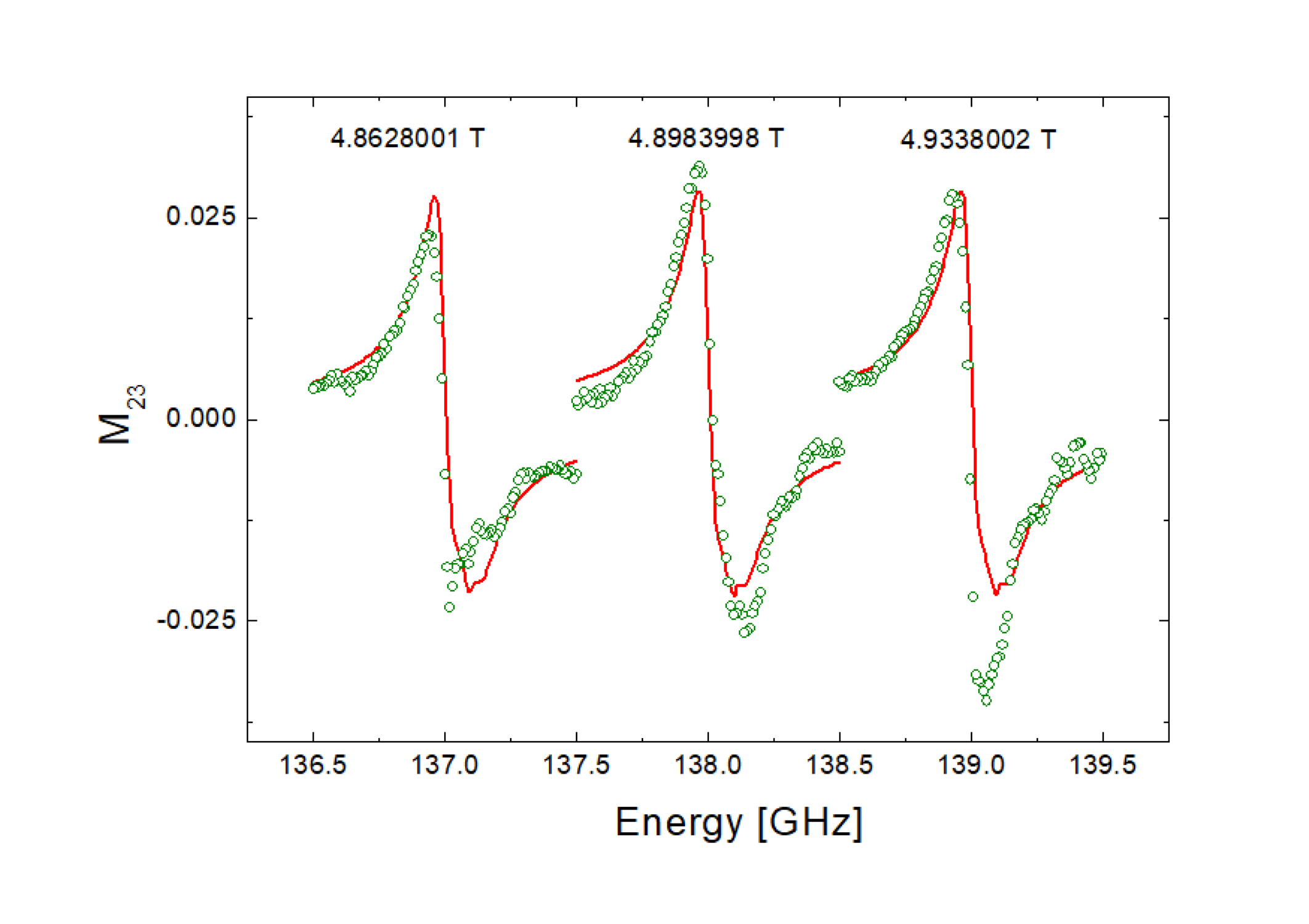}
\caption{Experimental (symbols) and best-match model calculated (solid lines) THz-EPR-GSE Mueller matrix data M$_{23}$ as a function of frequency for three different magnetic fields.}
\label{fig:SiCfreq}
\end{figure}

\begin{figure}[ht]
\centering
\includegraphics[width=1.00\linewidth]{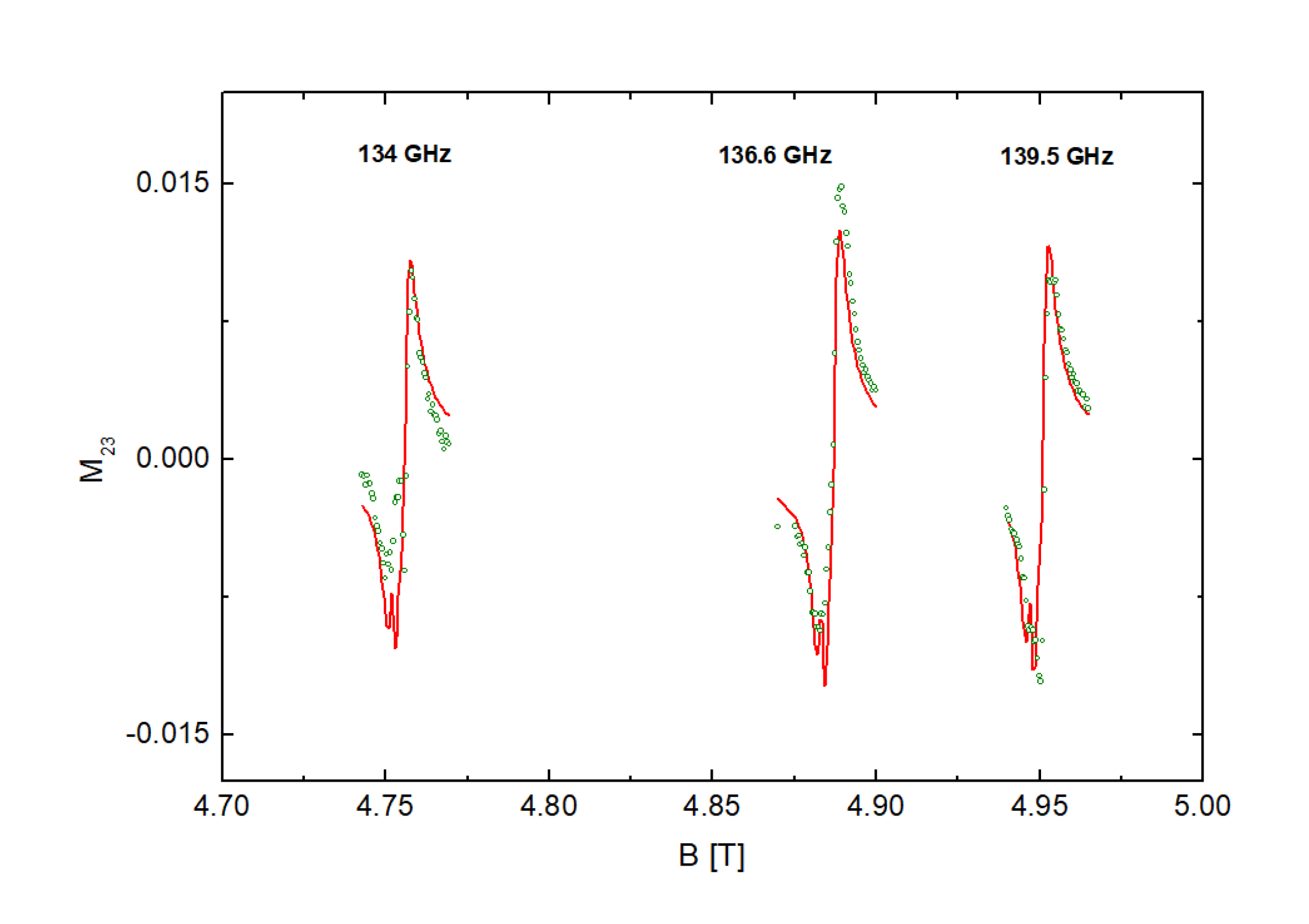}
\caption{Same as Fig.~\ref{fig:SiCfreq} as a function of magnetic field for various fixed frequencies.}
\label{fig:SiCfield}
\end{figure}

\begin{figure*}[ht]
\centering
\includegraphics[width=1.00\linewidth]{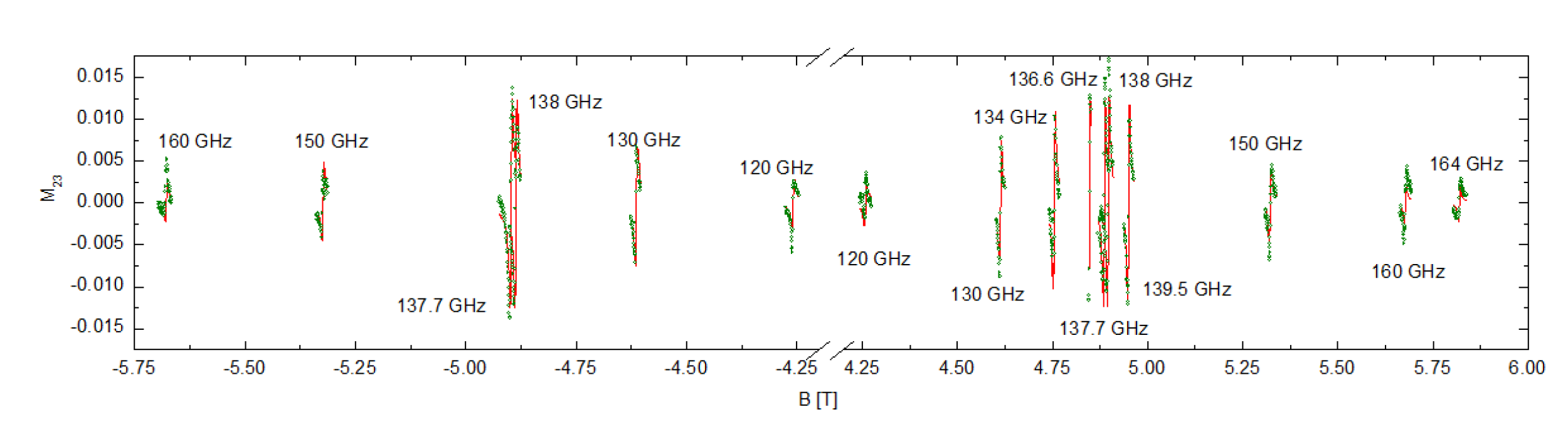}
\caption{Same as Fig.~\ref{fig:SiCfield}, including positive and negative fields.}
\label{fig:SicAllfield}
\end{figure*}

\begin{figure}[ht]
\centering
\includegraphics[width=1.00\linewidth]{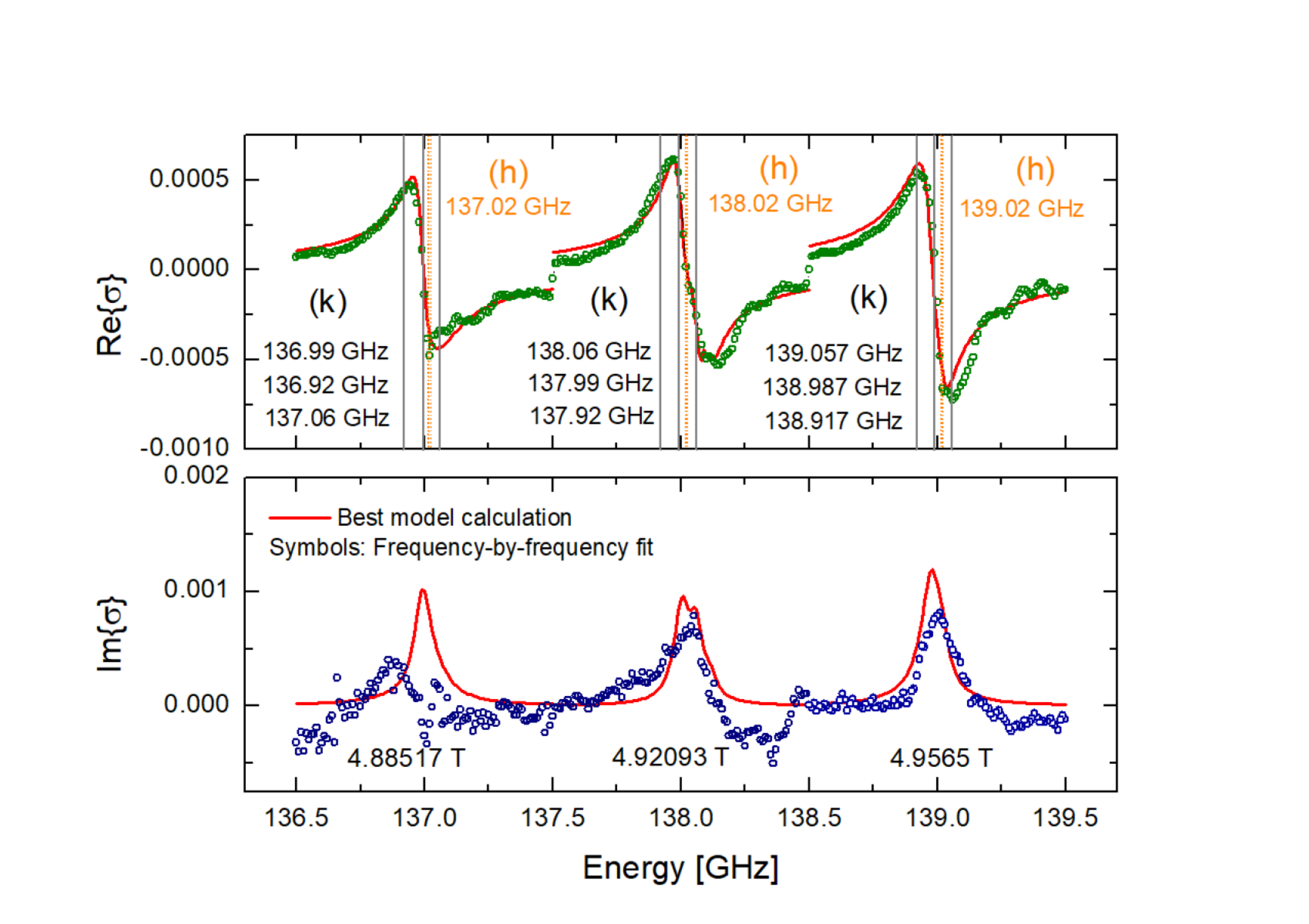}
\caption{Symbols: Real (top) and imaginary (bottom) frequency-by-frequency best-match model calculated magnetic function $\sigma$ defined in Eq.~\ref{eq:EPRtensorSiC}. Solid lines: Best-match model calculated lineshape functions defined in Eq.~\ref{eq:lorentz}. Vertical lines indicate frequencies listed for the $k$ and $h$ site triplets at the given magnetic field values, obtained from the lineshape functions and literature data as described in Table~\ref{Table:EPRparameters}.} 
\label{fig:Sigma}
\end{figure}


Figure~\ref{fig:SiCfreq} depicts selected THz-EPR-GSE data from the N-doped (0001) 4H-SiC sample across the frequency region of the $h$ and $k$ sites spin resonances at three different, fixed magnetic fields. Data shown here are differences between data at positive fields and at negative fields of the same magnitude. We only show here element $M_{23}$, which is equal to $M_{32}$, while all other elements are much smaller in magnitude. All Mueller matrix elements measured here are shown in the Supplementary material, as example for $|B| = 4.8984$~T. The signature in $M_{23}$ in Fig.~\ref{fig:SiCfreq} shifts linearly with magnetic field, in agreement with the linear Zeeman splitting of the $hfs$-split spin levels with magnetic field amplitude at very large frequencies. Solid lines depict best-match model calculated data. A small variation of the lineshape with field is observed in the experimental data, while the model calculated lineshape remains nearly the same. We believe this is due to magnet inhomogeneities and instrument imperfections at this point, see further discussion below. Figures~\ref{fig:SiCfield} and~\ref{fig:SicAllfield} depict selected data at fixed frequencies versus magnetic field. Of note is the reversal of the features in $M_{23}$ because now the field dependent resonance energy, $\Delta E(\mathbf{B})$, is the fixed term in the denominator of Eq.~\ref{eq:lorentz}. The signatures also change sign with field direction (Fig.~\ref{fig:SicAllfield}). Optical interference within the SiC substrate causes variation of the signature amplitude with frequency. 

The resonance features seen in Figs.~\ref{fig:SiCfreq},~\ref{fig:SiCfield},~\ref{fig:SicAllfield} can be well described by Eq.~\ref{eq:lorentz}. Two sets of three functions in Eq.~\ref{eq:EPRtensorSiC} are used, $\sigma_{(h,k),j}$, $j = 1,2,3$. The first set renders the triplet for $k$ site transitions, the second for $h$ site transitions. Amplitude ($A_{(h,k),j}$), broadening ($\gamma_{(h,k),j}$), $g_{h,k}$ and $\alpha_{(h,k),hfs}$ values can be treated as adjustable parameters. The best model calculations are shown in Figs.~\ref{fig:SiCfreq},~\ref{fig:SiCfield}, and~\ref{fig:SicAllfield} as solid lines. The main features are due to the $k$ transitions, which are broadened to the extend that individual transitions are smeared out. The $h$ transitions are rather small, and can only be resolved as a small deviation from the $k$ transition lineshapes. The line broadening, reflected in the broadening parameters $\gamma_{(h,k),j}$, is dominated by the magnetic field variation across the sample. Our magnet system (8~T magneto-optical cryostat, Cryogenic Ltd., London, U.K.)\cite{WOS:000431446900001} is not optimized for field homogeneity (field flattened), which is required for high-resolution EPR. We estimate an upper limit of ca. 0.15\% field deviation across the sample, e.g., ca. 7.8~mT at 5~T nominal field (See supplementary materials). Hence, our current THz-EPR-GSE data appear much more broadened than previous data obtained at, e.g., 142~GHz in differential-absorbance EPR within field-flattened magnets (e.g., Ref.~\onlinecite{GreulichWeber1997}). Nonetheless, our best model calculations still permit to identify the parameters listed in Table~\ref{Table:EPRparameters}, and which were obtained by simultaneously fitting all data sets, including all Mueller matrix elements differences, to one common set of values for the $h$ and $k$ site transitions. For the $k$ site transitions $g$ and $hfs$ values are in excellent agreement with previous results. Due to the broadening, parameters for the $h$ site transitions had to be assumed here. The supplementary material contains complete sets of Mueller matrix data for a field scan and for a frequency scan, as examples.

\begin{table*}\centering 
\caption{\label{Table:EPRparameters}Best-match and assumed model parameters for the THz-EPR-GSE model for the $h$ and $k$ site spin triplet transitions for the N$_C$ defect in 4H-SiC.}
\begin{ruledtabular}
\begin{tabular}{{l}{c}{c}{c}{c}{c}{c}{c}{c}{c}}
Transition& $A_{1}$ (mT)&$A_{2}$  (mT)&$A_{3}$  (mT)&$\gamma_{1}$  (mT)&$\gamma_{2}$  (mT)&$\gamma_{3}$  (mT)&$g$&$\alpha_{hfs}$ (mT)\\
\hline
$k$(j)& 0.75$\pm$0.01& 2.4$\pm$0.01& 3.4$\pm$0.01& 1$\pm$0.1& 2$\pm$0.1& 1$\pm$0.1& 2.0025$\pm$0.0001&2.0$\pm$0.1\\
 & & & & & & & 2.0028$^{a}$&1.8$^{a}$\\
$h$(j)& 0.01$\pm$0.01$^{b}$&0.01$\pm$0.01$^{b}$&0.01$\pm$0.01$^{b}$&0.01$\pm$0.01$^{b}$&0.01$\pm$0.01$^{b}$&0.01$\pm$0.01$^{b}$& $g_k + 0.00045^{c}$&0.10$^{d}$\\
 & & & & & & & 2.00325$^{b,d}$&0.10$^{b,d}$\\
\end{tabular}
\end{ruledtabular}
\begin{flushleft}
\footnotesize{$^\textrm{a}${Ref.~\onlinecite{GreulichWeber1997}}.}\\
\footnotesize{$^\textrm{b}${Parameters for transitions 1,2, and 3 assumed equal.}}\\
\footnotesize{$^\textrm{c}${Parameter shifted according to Ref.~\onlinecite{GreulichWeber1997}.}}\\
\footnotesize{$^\textrm{d}${Parameter not varied during model calculations.}}\\

\end{flushleft}
\end{table*}

Figure~\ref{fig:Sigma} depicts the real (top) and imaginary (bottom) frequency-by-frequency best-match model calculated magnetic susceptibility function $\sigma$ (Eq.~\ref{eq:EPRtensorSiC}; symbols). These data are  analogue to the complex-valued dielectric function, including anisotropy, determined by generalized spectroscopic ellipsometry. In the frequency-by-frequency data inversion approach, a best-match calculation is obtained by varying the parameters of $\sigma$ at every frequency independent of parameter values at any other frequency, i.e., without model lineshape assumption. Note also that there are three different spectra shown in Fig.~\ref{fig:Sigma}, each of which is fitted separately to the measurements shown in Fig.~\ref{fig:SiCfreq}. The solid lines represent the best-match model using Eq.~\ref{eq:lorentz}. Vertical lines indicate the spectral positions of the spin triplets. The main features are caused by strong amplitudes of LCP light absorption due to $k$ site triplet resonances. The $h$ site triplet is too small to be resolved. We also note that the real part is directly proportional to Mueller matrix elements $M_{23,32}$, while the imaginary parts are proportional to $M_{12,21}$ and  $M_{13,31}$ and which are much smaller and carry larger error bars in our measurements. Note that the real part of $\sigma$ is proportional to the circular birefringence while the imaginary part is proportional to the circular dichroism caused by the spin resonances. The latter part is primarily responsible for absorbance at normal incidence, while the former is much more sensitively determined in ellipsometry. A fact that is long known,\cite{Drude_1900} and which is reflected here again in a thin film optical situation, where the THz wavelength is much larger than the thin film (SiC substrate) thickness. Hence, we conclude that THz-EPR-GSE is a new method with high sensitivity to the  complex-valued magnetic response of thin film samples and by extension also in thin film heterostructures. The concept of generalized ellipsometry involved in THz-EPR-GSE provides promising prospects to characterize spin systems in highly anisotropic materials, for example in monoclinic symmetry $\beta$-Ga$_2$O$_3$.\cite{doi:10.1063/5.0031464} We note that knowledge of the real and imaginary spectra of the spin transitions should be sufficient to calculate the volume density of spins, similar to the dipole density derived from the complex valued dielectric function.\cite{Poole1983,WeilBolton2007}


We have demonstrated measurement of spin resonance at high magnetic fields using ellipsometry principles, and determined the complex valued magnetic polarizability of nitrogen doped 4H-SiC in the THz spectral range. Lineshape analysis using Lorentzian broadened harmonic oscillators provides excellent matches to our measured data, and permits identification of the $g$ factors and hyper fine splitting parameters, in principle. The ability to measure the frequency dependent complex-valued magnetic response function promises access to the polarization properties of defects in highly anisotropic materials. Future instrumentation improvement will enable higher magnetic field resolution, for example, by development of field flattened split coil magnet systems.


We acknowledge support by the National Science Foundation awards DMR 1808715, OIA-2044049, by Air Force Office of Scientific Research awards FA9550-18-1-0360, FA9550-19-S-0003, and FA9550-21-1-0259, by the University of Nebraska Foundation and the J.~A.~Woollam~Foundation, by the Swedish Research Council VR award No. 2016-00889, the Swedish Foundation for Strategic Research Grant Nos. RIF14-055 and EM16-0024, by the Swedish Governmental Agency for Innovation Systems VINNOVA under the Competence Center Program Grant No. 2016–05190, and by the Swedish Government Strategic Research Area in Materials Science on Functional Materials at Link{\"o}ping University, Faculty Grant SFO Mat LiU No. 2009-00971. This work is performed in the framework of the Knut and Alice Wallenbergs Foundation funded grant "Wide-bandgap semiconductors for next generation quantum components"   (Grant No. 2018.0071). M.~S. acknowledges stimulating discussions with Professor Patrick Dussault. We thank Professors Andreas P\"{o}ppl, Ivan Gueorguiev Ivanov, and Tien Son Nguyen for valuable comments and discussions. We thank Thorsten Maly (Bridge12) for help with magnet calibration.

The data that support the findings of this study are available from the corresponding author upon reasonable request.

\bibliography{main.bbl}
\end{document}